\documentclass[11pt,a4paper]{article}
\pdfoutput=1
\usepackage{geometry}
\usepackage[pdftex]{graphicx}
\usepackage[usenames, dvipsnames]{color}
\usepackage{framed}
\usepackage[bottom]{footmisc}
\usepackage{amssymb}
\usepackage{latexsym}
\usepackage{amsmath}
\usepackage{enumitem}
\usepackage{bm}
\usepackage{mathrsfs}

\newcommand{\dd}{\text{d}}
\newcommand{\V}{\mathcal{V}}
\newcommand{\hgh}{hGH}
\newcommand{\lif}{ R}
\renewcommand{\aa}{a}
\newcommand{\tr}{u}
\newcommand{\tth}{\chi}

\usepackage[retainorgcmds]{IEEEtrantools}
\usepackage[bookmarks=true,pdfborder={0 0 0}]{hyperref}
\usepackage[all]{hypcap}

\newtheorem{theorem}{Theorem}
\begin{document}

\title{Monopoles, instantons and the Helmholtz equation}
\author{Guido Franchetti\footnote{\tt{franchetti@math.uni-hannover.de}} \ and Rafael Maldonado\footnote{{\tt rafael.maldonado@itp.uni-hannover.de}}\vspace{0.2cm}\\
\emph{Institut f\"ur Theoretische Physik, Leibniz Universit\"at Hannover}\\\emph{Appelstra\ss e 2, 30167 Hannover, Germany}}

\maketitle
\vspace{-8cm}
\begin{minipage}{\linewidth}
\begin{flushright}
ITP-UH-07/16
\end{flushright}
\end{minipage}
\vspace{7cm}
\begin{abstract}
\noindent In this work we study the dimensional reduction of smooth circle invariant Yang-Mills instantons 
defined on 4-manifolds which are non-trivial circle fibrations over hyperbolic 3-space.  A suitable choice of the 4-manifold 
metric within a specific conformal class gives rise to singular and smooth hyperbolic monopoles.  A large class of monopoles is obtained if the conformal factor satisfies the 
Helmholtz equation on hyperbolic 3-space.  We describe simple configurations and relate our results to the JNR construction, for 
which we provide a geometric interpretation.

\end{abstract}

\section{Introduction}

\noindent Yang-Mills-Higgs monopoles over Euclidean $3$-space have received considerable attention both because of their rich 
mathematical structure \cite{AH88} and due to their connections to string theory, see e.g.~\cite{HW97}.  Qualitatively 
similar solutions are obtained by working on hyperbolic space, with the advantage that the monopole equations are then more 
readily integrated analytically, even for quite complicated configurations \cite{MS14}. 
The reason for this simplification is that hyperbolic monopoles with specific masses arise as symmetry reductions of smooth 
circle invariant Yang-Mills instantons on Euclidean $4$-space \cite{Ati88}.

Euclidean monopoles with point singularities were first studied by Kronheimer \cite{Kro} and have been constructed via the Nahm 
transform \cite{BC11}.  The analysis was extended to the hyperbolic case by Nash \cite{Nas08}, who studied the charge $1$ 
moduli space using twistor techniques.

In this paper we present a technique which allows one to construct both smooth and singular hyperbolic monopoles of mass 
$m=\tfrac{1}{2}$.  We make use of the fact that hyperbolic monopoles can be obtained from smooth circle invariant instantons via dimensional 
reduction.  This is  true not only for smooth monopoles, but also for monopoles with singularities. Smooth monopoles come from 
instantons living on Euclidean 4-space $E ^4 $, which is conformally equivalent to a trivial circle bundle over hyperbolic 3-space 
$H ^3 $. Singular monopoles, on the other hand, arise from the dimensional reduction of smooth instantons living on non-trivial 
circle fibrations over $H ^3 $. These spaces are the hyperbolic version of  Gibbons-Hawking gravitational instantons and we 
review them in Section \ref{hghs}. Sections \ref{shm} and \ref{dr} review some material on singular hyperbolic monopoles and dimensional reduction.

In Section \ref{hyphel} we construct monopoles by making use of the fact that the projection of the spin 
connection of a Riemannian 4-manifold $M$ on the appropriate $ \mathfrak{ su}(2) $ factor in the Lie algebra decomposition 
$ \mathfrak{ so} (4) \simeq \mathfrak{ su }(2) \oplus \mathfrak{ su }(2) $  is an instanton provided that $M$ is spin, half 
conformally flat and scalar flat \cite{AHS78}. In particular, by rescaling a hyperbolic Gibbons-Hawking space via a conformal factor which is circle invariant and satisfies the Helmholtz 
equation on $H ^3$, we obtain a large family of solutions of the Bogomolny equations. Some specific examples are studied in 
Sections \ref{superposition} and \ref{Backlund}. In Section \ref{hmm}  we remark on how this technique can be modified to 
obtain the Higgs field of spherically symmetric monopoles with higher mass. A novel family 
 of monopoles is obtained by imposing that the conformally rescaled manifold is Einstein. We study this family in Section 
\ref{emono}.

Reexamining the case of circle invariant instantons on $E ^4  $ from this perspective gives new insight into the construction of 
hyperbolic monopoles from circle invariant JNR data.  The JNR method constructs a class of hyperbolic monopoles starting from 
circle invariant harmonic functions on $E^4$ \cite{JNR77,BCS15}.  In Section \ref{jne} we show that  smooth monopoles obtained from solutions of the 
Helmholtz equation correspond to monopoles coming from JNR data, and we thereby obtain a purely geometrical reformulation of the 
JNR construction applied to hyperbolic monopoles.  Moreover, we identify how the classical JNR data is 
modified to generate monopole singularities and give a physical interpretation of the JNR poles.  We clarify the relation between the two approaches and how to translate from one to the other.

A property of hyperbolic monopoles which is not shared by their Euclidean counterparts is the fact that they are completely 
determined by the induced asymptotic Abelian connection \cite{BA90,MNS03}.  For monopoles coming from solutions of the Helmholtz 
equation we  show this directly in Section \ref{bdata} by giving an explicit relation between the full monopole solution and its asymptotic data.

\section{Preliminaries}\label{prelims}
 It is known \cite{Kro} that a circle invariant instanton on a Gibbons-Hawking space is equivalent to a monopole on 
 $E^3$, possibly with singularities, and examples of these monopoles have been constructed \cite{EH03}.  In a similar way, it is 
 possible to obtain a singular hyperbolic monopole starting from a circle invariant instanton on  modified Gibbons-Hawking spaces \cite{Nas08}.  We are now going to review how to generate such instantons.

\subsection{Hyperbolic Gibbons-Hawking spaces} 
\label{hghs} 
A (Euclidean) Gibbons-Hawking space is a Riemannian $4$-manifold $M$ with a metric of the form
 \begin{equation}
 \label{GH} 
  \V g_{E^3}+\V^{-1}(\dd\psi+\aa)^2.
 \end{equation}
 Here $g_{ E^3}$ is the Euclidean $3$-metric and $(\V,\aa)$ obeys the Abelian monopole equation 
 $\dd\aa=\ast_{ E^3}\dd\V$, hence $\V$ is harmonic on $ E^3$.  Any such metric is hyper-K\"ahler 
 and therefore Ricci-flat and half-conformally flat.  Let $G^{E^3}_{p_i} $ be the Green's function centered at $p_i\in E^3$, $G^{E^3}_{p_i} (p) = \mu / d ^E (p _i ,p)$, with $d ^E $ the Euclidean distance in $E ^3 $ and $\mu>0$ a constant related to the range of the angle $\psi$ by $\psi \in[0, 4 \pi \mu )$. For $\V$ of the form
 \begin{equation}
  \V\,\propto\,c+\sum_iG^{E^3}_{p_i},\qquad\qquad(c\,=\,\text{constant})
 \end{equation}
the metric \eqref{GH} is known as multi-Eguchi-Hanson if 
$c=0$ and as multi-Taub-NUT otherwise.  It has a $U(1)$ isometry group generated by the vector field 
$\partial/\partial\psi$.  Away from the NUTs $p_i$, the fixed points of the $U(1)$ action, $M$ is the total space of a circle 
fibration over $ E^3$.

%\red{We need to explain that $\psi$ has range $[0,2\pi)$ if $c\neq0$ and range $[0,4\pi)$ if $c=0$.  One paper one could cite is \cite{GH79}.}

LeBrun \cite{LeB91} obtained a new family of half-conformally flat spaces by replacing the base space $E^3$ with hyperbolic 
$3$-space $H ^3 $.  The metric is now
 \begin{equation}
\label{HGH}
  g_0\,=\,V g_{H^3}+V^{-1}(\dd\psi+\alpha)^2,
 \end{equation}
where $g_{H^3}$ is the metric on hyperbolic $3$-space of sectional curvature $-1$ and $(V,\alpha)$ is an Abelian monopole on 
$H^3$.  We will refer to these spaces as hyperbolic Gibbons-Hawking (\hgh) spaces and take the orientation specified by the volume form
\begin{equation}
\mathrm{vol} _{ \hgh} = - V \, \mathrm{vol} _{  H ^3 }\wedge \mathrm{d} \psi ,
\end{equation} 
where $ \mathrm{vol} _{ H ^3 }$ is the volume element on $H ^3 $.  Note that  \eqref{HGH} is neither hyper-K\"ahler nor scalar 
flat.  In fact \cite{LNN97}, the scalar curvature $s$ is
\begin{equation}
\label{sss} 
 s\,=\,-\frac{6}{V}.
\end{equation}
We will take $V$ to be of the form
\begin{equation}
\label{vvv} 
 V\,=\,\frac{2}{\beta}+2\sum_iG^{H^3}_{p_i},\qquad\qquad(\beta\,=\,\text{constant})
\end{equation}
where $G^{H^3}_{p_i}$ is the Green's function centered at $p_i\in H^3$.  If $d ^H $ is the distance function on $H^3$, then
\begin{equation}
 G^{H^3}_{p_i}(p)\,=\,\frac{1}{\text{e}^{2d ^H (p,p_i)}-1}.
\end{equation}
With this  normalisation of $V$, the range of $\psi$ in (\ref{HGH})  is $ \psi \in [0, 4 \pi )$.

\subsection{Singular hyperbolic monopoles}
\label{shm} 
A hyperbolic monopole $(\Phi, \mathcal{A} )$  is a solution of the Bogomolny equations on hyperbolic space,
\begin{equation}
\label{bogo} 
\mathrm{d} _{ \mathcal{A} } \Phi =  - * _{ H ^3 } \mathcal{F},
\end{equation} 
where $\mathcal{A} $ is a connection on an $SU (2) $ bundle $\mathcal{P} $ over $H ^3 $, $\mathcal{F} $ is its curvature, $\Phi$ is a section of the adjoint bundle and $* _{ H ^3 }$ denotes the Hodge operator on $H ^3 $. On $\mathfrak{ su }(2)$ we take  the ad-invariant inner product  $\langle X,Y \rangle   = - \tfrac{1}{2} \mathrm{Tr} ( X \, Y )$.

  Let $B ^{\parallel}$, $B ^{ \perp}$  be the  components of $\mathcal{A} $ parallel and orthogonal to the direction of the 
  Higgs field in $\mathfrak{su}(2)$,  $B ^{ \parallel} = \langle \mathcal{A} , \Phi  \rangle\, \Phi /\| \Phi \| ^2  $, $B ^{ \perp}= \mathcal{A} - B ^{ \parallel} $.
The monopole is required  to satisfy the following asymptotic conditions \cite{MS00,Nas08}:
\begin{align} 
\label{as1} 
&(\|\Phi \| -m  ) \exp(2 r) \text{ extends smoothly to $\partial H ^3 $, for some constant $m >0$},\\
\label{as2} 
& B ^{ \parallel}  \text{ extends smoothly to $\partial H ^3 $},\\
\label{as3} 
&B ^{ \perp} \exp(2m \, r)  \text{ extends smoothly to $\partial H ^3 $}.
\end{align}

Let $\{ p _i \} $ be distinct points in $H ^3 $, $r _i = d ^H (p,p _i ) $.
A singular hyperbolic monopole with singularities at  $\{p _i\} $ is a solution of (\ref{bogo}) on $H ^3 \setminus \{ p _i \} $ which satisfies the following conditions:
\begin{align}
\label{s1} 
&2\lim _{ r _i \rightarrow 0 } r _i  \| \Phi \| = \ell  _i \in \mathbb{Z}  _+ ,\\
\label{s2} 
& \mathrm{d} (r _i \| \Phi \| )  \text{  is bounded in a neighbourhood of $p _i $}.
\end{align} 

The quantity
\begin{equation}
\ell =\sum _i \ell _i 
\end{equation} 
is called the Abelian charge of the monopole.
The total charge $Q$ of a hyperbolic monopole is the first Chern number of the asymptotic Abelian  fibration. It can be computed as
\begin{equation}
\label{qqqq} 
Q 
= -  \frac{1}{2 \pi } \int _{ \partial H  ^3 } * _{ H ^3 } \mathrm{d} \| \Phi \|.
\end{equation} 
It follows from (\ref{as1}) that the coefficient of the leading term in an asymptotic expansion of $ \| \Phi \|  $  is the monopole mass $m$. If $C$ is the smooth extension of $(\|\Phi \| -m  ) \exp(2 r) $ to $\partial H ^3 $, then
\begin{equation}
Q =
\frac{1}{4 \pi }\int _{ \partial H ^3 }C \,  \mathrm{vol} _{ S ^2 } .
\end{equation}
Followingg \cite{Nas08}, we define the non-Abelian charge $k$ to be
\begin{equation}
\label{klq}
k = \ell - Q.
\end{equation}

\subsection{Dimensional reduction}
\label{dr} 
Let $P$ be the total space of an $SU (2) $ principal bundle over an \hgh~space $M$.
An $SU(2)$ instanton on $M$  is a connection on $P$ having self-dual or anti-self-dual curvature. 

Let  $\lif $ be a lift to $P$ of the $U (1)  $ action on $M$ generated by the Killing vector field $\partial / \partial \psi $. We say that an instanton $\varpi $ on $M$ is $\psi$-invariant if $ \lif ^\ast \varpi = \varpi $. For a $\psi$-invariant instanton  $\varpi$ there exists a local section $s$ such that, away from fixed points of the $\partial / \partial \psi $ action,  the gauge potential $A =s ^\ast \varpi $ has no explicit $\psi$ dependence \cite{MW96}. We call such a choice of local section a circle invariant gauge.

A $\psi$-invariant self-dual instanton is equivalent to a hyperbolic monopole. In fact \cite{Kro,Nas08}, working in a circle invariant gauge $s$, and writing
\begin{equation}
A \, = \, s ^\ast \varpi \, =\, \mathcal{A} + \frac{ \Phi } {V } (\mathrm{d} \psi + \alpha ),
\end{equation} 
 the self-duality equations for $A $ imply the Bogomolny equations (\ref{bogo}) for $(\Phi, \mathcal{A}) $.

One could reverse the procedure and define an instanton $A$ on an \hgh~space $M$ from a singular hyperbolic monopole. Conditions 
(\ref{s1}), (\ref{s2}) then ensure that $A$ is globally defined on $M$ as long as the monopole singularities are located 
at poles of $V$ \cite{Kro,Nas86}.

\section{The conformal rescaling method}\label{themethod}
For an oriented spin 4-manifold there is a decomposition of the spin bundle $ S (M) = S ^+ (M) \oplus  S ^{-} (M) $ corresponding to the splitting $ \mathfrak{  so }(4) \simeq \mathfrak{ su }(2) ^+ \oplus \mathfrak{ su }(2) ^- $. Denote by  $P _{ \pm }( \omega) $ the projection of the spin connection $\omega$  onto $S ^{\pm} (M) $.
We will make use of the following result \cite{AHS78}.
\begin{theorem}[Atiyah-Hitchin-Singer 1978]
\label{ahs} 
Let $M$ be an oriented Riemannian spin 4-manifold with spin connection $\omega$. 
\begin{enumerate}
\item  $P _- (\omega )$ is a self-dual $SU (2) $  connection  if and only if $M$ is half conformally flat and scalar flat.
\item   $P _+ ( \omega ) $ is a self-dual $SU (2) $ connection if and only if $M$ is Einstein.
\end{enumerate} 
\end{theorem} 

Since the self-duality equations are conformally invariant, we can conformally rescale the metric on an \hgh~space in order to get a metric satisfying either of the above conditions. The appropriate projection of the spin connection is then a self-dual instanton on the \hgh~space. In the Euclidean case, this method has been used e.g.~in \cite{EH03}.

\subsection{Hyperbolic monopoles as solutions of the Helmholtz equation}
\label{hyphel} 
In this section we are going to apply the first method of theorem \ref{ahs} to generate self-dual instantons on \hgh~spaces. We shall see that they can be completely specified by giving a solution of the Helmholtz equation.
%\begin{equation}
%\triangle _{ H ^3 } \Lambda + \Lambda =0,
%\end{equation} 
%where $ \triangle _{ H ^3 }$ is the Laplace operator on $H ^3 $.

Since half conformal flatness is a conformally invariant condition, we are looking for a function $\Lambda >0$ for which  the metric 
\begin{equation}
\label{gcf} 
g 
\, =\,  \Lambda ^2 \, g _0 
\,=\, \Lambda  ^2 \left[ V g_{H^3}+V^{-1}(\dd\psi+\alpha)^2 \right]
\end{equation} 
is scalar flat. Under the conformal transformation (\ref{gcf}), the scalar curvature $s$ of  $g _0 $ transforms as
\begin{equation}
s ^\prime = \frac{1}{\Lambda ^2 } \left( s - 6\, \frac{ \triangle _{ g _0 } \Lambda }{\Lambda } \right).
\end{equation} 
We use the notation $\triangle _{ g } $ to denote the Laplace-Beltrami operator with respect to the metric $g $.
Let us assume that $\partial _\psi \Lambda =0 $.
Then $ \triangle _{ g _0 } =  V ^{-1} \triangle _{  H ^3  }$ and, using (\ref{sss}),  imposing $s ^\prime =0$ gives the Helmholtz equation on hyperbolic space,
\begin{equation}
\label{hel} 
\triangle _{ H ^3 }\Lambda  + \Lambda \, = \, 0.
\end{equation} 

As we are going to show below, in terms of $\Lambda$ the instanton obtained from the spin connection $\omega $ on the conformally \hgh~space with metric (\ref{gcf})   is given by
\begin{equation}
\label{inst} 
A = \mathcal{A} + \frac{ \Phi }{V } ( \mathrm{d} \psi + \alpha ),
\end{equation} 
with $(\Phi, \mathcal{A}  )$ the hyperbolic monopole
\begin{align}
\label{toph} 
\Phi &
=  \frac{1}{2}  \left[ 
\frac{\epsilon _1  (\Lambda) }{\Lambda } \mathbf{i}
+ \frac{\epsilon _2 (\Lambda) }{\Lambda } \mathbf{j}
+ \frac{\epsilon _3 (\Lambda) }{\Lambda }\, \mathbf{k} 
  \right] ,\\
\label{topgp} 
\mathcal{A} &
= \frac{1}{4}  \left[
( \mathcal{C} _{3k2} + \mathcal{C} _{k23} - \mathcal{C} _{ 23k } )\,  \mathbf{i} +
( \mathcal{C} _{1k3} + \mathcal{C} _{k31} - \mathcal{C} _{ 31k } )\,  \mathbf{j} +
(\mathcal{C} _{2k1} + \mathcal{C} _{k12} -\mathcal{C} _{ 12k } )\,  \mathbf{k} 
\right] \epsilon ^k  + \\ \nonumber& 
+ \frac{1}{2 \Lambda } \left[ 
\left(\epsilon _2 (\Lambda )\epsilon ^3- \epsilon _3 (\Lambda )\epsilon ^2     \right) \mathbf{i} + 
\left(\epsilon _3 (\Lambda )\epsilon ^1 - \epsilon _1 (\Lambda )\epsilon ^3     \right) \mathbf{j} + 
\left( \epsilon _1 (\Lambda )\epsilon ^2 - \epsilon _2 (\Lambda )\epsilon ^1   \right) \mathbf{k}
\right] .
\end{align} 
Here $\{ \epsilon _i \} $ is an orthonormal frame on $H ^3 $, $\{ \epsilon  ^i \} $ is the dual coframe,
$
[ \epsilon _i , \epsilon _j ] =\mathcal{C} _{ij} ^{ \phantom{ ij }k } \epsilon _k 
$
and $\mathcal{C} _{ ijk} =\mathcal{C} _{ij} ^{ \phantom{ ij }m } \, (g _{ H ^3 }) _{ mk } $. Latin indices range from 1 to 3.
%\red{NOTE}: we are using $ \mathbf{i} ^\prime = - \mathbf{k} $, $ \mathbf{j} ^\prime =- \mathbf{j} $, $ \mathbf{k} ^\prime =- \mathbf{i} $.

To prove (\ref{toph}), (\ref{topgp})  take the $g$-orthonormal coframe $\{ e ^ i, e ^4   \} $ and dual frame  $\{ e  _i , e _4   \} $,
\begin{IEEEeqnarray}{rclcrcl}
\label{ocof} 
 e ^i &\, =\, &\Lambda \sqrt{ V } \, \epsilon ^i,&
 \qquad  &e ^4 &\, =\, & - \frac{\Lambda\,  (\mathrm{d} \psi + \alpha) }{\sqrt{ V }},\\ \nonumber
e _i &\, =\, &\frac{\epsilon _i }{\Lambda \sqrt{ V }},& \qquad &
e _4 &\, =\, & - \frac{\sqrt{  V }  }{\Lambda }\, \partial _\psi. 
\end{IEEEeqnarray} 
Let  $[e _ \alpha  , e _ \beta  ]= C _{\alpha \beta  }^{ \phantom{ \alpha \beta  } \gamma  } e _ \gamma  $, $C _{ \alpha \beta \gamma }= g _{ \mu \gamma }  C _{ \alpha \beta }^{ \phantom{ \alpha \beta  }\mu  }$, with Greek indices ranging from 1 to 4.
The spin connection coefficients can be computed making use of the equation
\begin{equation}
\omega _{ \alpha \beta }
= \frac{1}{2}  \left( C _{ \alpha \beta \mu } - C _{ \beta \mu \alpha  } - C _{ \mu \alpha \beta } \right) e ^\mu .
\end{equation} 
Since
\begin{equation}
\begin{split} 
C _{ i4 }^{ \phantom{ i4 }4 }&
= \frac{\mathrm{d} V (\epsilon  _i )}{2\Lambda  V ^{ 3/2 } } - \frac{\mathrm{d} \Lambda (\epsilon  _i ) }{\Lambda^2 V ^{ 1/2 } },\\
C _{ ij }^{ \phantom{ ij }4 }&
= \frac{* _{ H ^3 }\mathrm{d} V (\epsilon  _i , \epsilon  _j )}{\Lambda V ^{ 3/2 }} ,\\
 C_{ i 4} ^{ \phantom{ ik }j } &=0,\\
 C _{ ij }^{ \phantom{ ij }k }&
=\frac{1}{\Lambda  \sqrt{ V } } \left(\mathcal{C}  _{ ij }^{ \phantom{ ij }k} + \epsilon _j  (\sqrt{ V }\Lambda )\, \delta _i ^k  
- \epsilon _i  (\sqrt{ V }\Lambda )\, \delta _j ^k   \right),
\end{split} 
\end{equation} 
we have,
\begin{equation}
\begin{split} 
\omega _{ ij }&
=\frac{1}{2 } \left( \mathcal{C} _{ ijk }- \mathcal{C} _{ jki }- \mathcal{C} _{ kij } \right) \, \epsilon ^k 
+\frac{1}{\Lambda } \left[ \epsilon _j (\Lambda) \, \epsilon ^i - \epsilon _i (\Lambda) \, \epsilon ^j  \right] 
+ \frac{1}{2 V } \left[ \epsilon _j (V) \, \epsilon ^i - \epsilon _i (V) \, \epsilon ^j  \right]  \\ &
- \frac{ * _{ H ^3 } \mathrm{d} V (\epsilon _i , \epsilon _j )}{2 V }\left(\frac{\mathrm{d} \psi + \alpha }{V }\right) ,\\
\omega _{ i 4 } &
=\left( \frac{\mathrm{d} \Lambda (\epsilon _i ) }{\Lambda } - \frac{\mathrm{d} V (\epsilon _i ) }{2 V } \right)
\left(  \frac{\mathrm{d} \psi + \alpha }{V } \right) 
+ \frac{ * _{ H ^3 } \mathrm{d} V (\epsilon _i , \epsilon _j )}{ 2 V } \, \epsilon ^j .
\end{split} 
\end{equation}

The projection of $\omega$ onto $S ^- (M) $ is given by
\begin{equation}
P _{ - }(\omega)  
=\frac{1}{4} (\bar \eta _a ) _{ i  j} \, \omega _{ i j} \, \bar \eta _a,
\end{equation} 
with $\{\bar \eta _i \}  $  the anti-self-dual 't Hooft matrices
\begin{equation}
\label{asdth}
\bar \eta _1 
= \begin{pmatrix} 0 &1 &0 &0\\ -1 &0 &0 &0\\0 &0 &0 &-1 \\ 0 &0 &1 &0\end{pmatrix}, \quad 
\bar \eta _2 
= \begin{pmatrix}
0 &0 &-1 &0\\ 0 &0&0 &-1\\ 1 &0 &0 &0\\ 0 &1 &0 &0
\end{pmatrix},\quad 
\bar \eta _3 =\begin{pmatrix}
0 &0&0&-1\\0&0&1&0\\0&-1&0&0\\1&0&0&0
\end{pmatrix}.
\end{equation} 
Identifying
\begin{equation}
\bar \eta _1 = - \mathbf{k} , \quad 
\bar \eta  _2 = - \mathbf{j} , \quad 
\bar \eta  _3 = - \mathbf{i} ,
\end{equation} 
we get
\begin{equation}
\label{nmsdghasf} 
A 
=P _{ - }(\omega)
= \frac{1}{2} \Big[ 
(\omega _{ 14 }- \omega _{ 23 } )\mathbf{i} + 
 (\omega _{ 24 }+ \omega _{13 }) \mathbf{j} + 
(\omega _{ 34 }- \omega _{ 12 } )\mathbf{k} 
 \Big].
\end{equation} 
Using
\begin{equation}
* _{ H ^3 } \mathrm{d} V (\epsilon _i , \epsilon _j )=\epsilon _{ ijk }\, \mathrm{d} V (\epsilon _k ),
\end{equation}
with $\epsilon _{ ijk }$ the Levi-Civita symbol in 3 dimensions, $\epsilon _{ 123 } =1$, we have
\begin{equation}
\begin{split} 
\omega _{ ij }- \epsilon _{ ijk }\, \omega _{ k 4 } &
= \frac{1}{2 } \left(  \mathcal{C} _{ ijk } - \mathcal{C} _{ jki } - \mathcal{C} _{ kij } \right) \, \epsilon ^k 
+ \frac{1}{\Lambda } \left[ \epsilon _j (\Lambda ) \, \epsilon ^i  - \epsilon _i (\Lambda ) \, \epsilon ^j   \right] \\ &
+\epsilon _{ ijk } \,  \frac{\mathrm{d} \Lambda (\epsilon _k ) }{\Lambda } \left(  \frac{\mathrm{d} \psi + \alpha }{V } \right) .
\end{split} 
\end{equation} 
By comparing (\ref{nmsdghasf}) with (\ref{inst}), equations (\ref{toph}), (\ref{topgp}) follow.

To see what constraints on $\Lambda$ result from  the asymptotic conditions (\ref{as1}) -- (\ref{as3}), let us take the hyperboloid model of $ H ^3 $ with metric
\begin{equation}
g _{ H ^3 } = \mathrm{d} r ^2 + \sinh ^2 r(\mathrm{d}\theta^2+\sin^2\theta\,\dd\phi^2).
\end{equation} 
The orthonormal coframe
\begin{equation}
\epsilon ^1 =\mathrm{d} r, \quad 
\epsilon ^2 =\sinh r \, \mathrm{d} \theta , \quad
\epsilon ^3 = \sinh r \sin \theta \, \mathrm{d} \phi,
\end{equation} 
has non-trivial commutation coefficients
\begin{equation}
\mathcal{C} _{ 122} =\mathcal{C} _{ 133 }=- \coth r, \qquad
\mathcal{C} _{ 233 } =- \frac{\cot \theta }{\sinh r} .
\end{equation} 
If $\Lambda$ satisfies the Helmholtz equation (\ref{hel}), which in these coordinates reads
\begin{equation}
\label{helhyp} 
\sinh ^2 r \left( \partial ^2 _r \Lambda +2 \operatorname{coth} r\,  \partial _r \Lambda + \Lambda \right) 
+ \partial ^2 _\theta \Lambda + \cot \theta\,  \partial _\theta \Lambda + \frac{1}{\sin ^2 \theta } \, \partial ^2 _\phi \Lambda  =0,
\end{equation} 
 equations (\ref{toph}), (\ref{topgp})  give the following solution of the Bogomolny equations
\begin{IEEEeqnarray}{rcl}
\Phi \,&
=&\,  \frac{1}{2\Lambda } \left(\partial _r \Lambda  \, \mathbf{i} + \frac{\partial _\theta  \Lambda  }{\sinh r}\, \mathbf{j}  +  \frac{\partial _\phi \Lambda }{\sinh r \, \sin \theta }\, \mathbf{k} \right) ,\label{phia}\\
\mathcal{A}\, &
=&
\,\frac{\mathbf{i} }{2} \left[ \left( \cos \theta  + \sin \theta\,  \frac{ \partial _\theta \Lambda }{ \Lambda }\right)  \, \mathrm{d} \phi - \frac{\partial _\phi \Lambda }{ \Lambda \sin \theta }\, \mathrm{d} \theta  \right]\label{phia2}\\
&-&\, \frac{\mathbf{j} }{2} \left[ \left(  \cosh r + \sinh r\,  \frac{ \partial _r \Lambda} {\Lambda }\right)  \sin \theta \, \mathrm{d} \phi - \frac{ \partial _\phi \Lambda }{ \Lambda \sinh r \sin \theta  }\, \mathrm{d} r \right]\nonumber\\
&+&\, \frac{\mathbf{k} }{2} \left[ \left(  \cosh r + \sinh r\,  \frac{ \partial _r \Lambda }{\Lambda } \right)   \, \mathrm{d} \theta - \frac{ \partial _\theta \Lambda }{\Lambda  \sinh r } \, \mathrm{d} r\right].\nonumber
\end{IEEEeqnarray} 
%Imposing (\ref{as1}) results in the condition
%$|\log \Lambda | = 2mr + \mathcal{O} (\exp(-2r) )$,
%while from (\ref{as2}), (\ref{as3})  we get $\log \Lambda = - r + \mathcal{O} (\exp(-2r) )$. Hence for these solutions
%\begin{equation} 
%m =1/2,
%\end{equation} 
%\begin{equation}
%\| \Phi  \| =  \frac{1}{2}  + Q \exp(-2r)
%\end{equation} 
Note that 
\begin{equation} 
\label{philambda}
\| \Phi \| ^2 
=- \frac{1}{4} \left(  1 + \triangle _{ H ^3  } \log \Lambda \right) 
= \frac{1}{4} | \mathrm{d} \log\Lambda | _{ H ^3 } ^2 .
\end{equation}

From the boundary conditions \eqref{as1} -- \eqref{as3} we find $m=\tfrac{1}{2}$ and
\begin{equation}
\label{asb} 
 \log\Lambda\,=\,-r+f_0 (\theta,\phi)+ f _2 (\theta,\phi)\, \text{e}^{-2r} + \mathcal{O} \left( \mathrm{e}^{ -4r }\right)   .
\end{equation}
It follows that
\begin{equation}
\label{fifi} 
 \|\Phi\|\,=\,\frac{1}{2}-\text{e}^{-2r}(\triangle_{S^2}f _0 -1)+\mathcal{O} (\text{e}^{-4r}).
\end{equation}
The total charge (\ref{qqqq})  corresponding to (\ref{fifi}) is then 
\begin{equation}
Q 
= 1- \frac{1}{4 \pi } \int _{ S ^2 } \triangle  _{ S ^2 }  f _0\,  \sin \theta \, \mathrm{d} \theta \wedge \mathrm{d} \phi 
=1 - \frac{1}{4 \pi } \int _{ S ^2 } \mathrm{d} * _{ S ^2 } \mathrm{d} f _0 .
\end{equation} 
In order for $Q$  to be an integer, we need
\begin{equation}
\label{akakaka} 
f _0  (\theta , \phi )
=-n \log (\sin \theta ) + \phi _0 ( \theta , \phi ),
\end{equation} 
with $n\in \mathbb{Z}  $ and $ \phi _0 $ such that $\mathrm{d} *_{  S ^2 } \mathrm{d} \phi  _0 $ is the zero element in 
$ H ^2 _{ \mathrm{dR}   } (S ^2 ) $, the second de Rham cohomology group of $S ^2 $. Hence   $ \mathrm{d} *_{  S ^2 } \mathrm{d} \phi  _0  =\mathrm{d} \beta $, 
$\beta \in \Omega ^1 (S ^2 ) $. Note that generally $ \beta \neq *_{  S ^2 } \mathrm{d} \phi _0 $ as 
$* _{ S ^2 } \mathrm{d} \phi _0 $ is not necessarily a globally defined 1-form on $S ^2 $. For $f _0 $ given by (\ref{akakaka}) we have $Q = 1-n $, while the monopole moduli are encoded in  $\phi _0 $.

Conditions (\ref{s1}), (\ref{s2})  imply that
\begin{equation}
\Lambda = r ^{ -\ell } + \mathcal{O} (r ^{ 1- \ell}) 
\end{equation} 
near a monopole singularity at $r =0$, and the Helmholtz equation requires us to take $\ell=1$.

 We would like to note here some properties of the circle invariant instanton (\ref{inst}).
 Let $\{ q _i \} $ be the set of singular points of $\Phi $,  $M$ be the conformally \hgh~space with metric (\ref{gcf}), and $\{ p _i \} $ the set of poles of $V$. 
First,  if equations (\ref{s1}) and (\ref{s2}) hold, then the instanton with gauge potential (\ref{inst}), a priori only defined on $M \setminus \{ p _i \} $,  actually extends to $M$ provided that $\{ q _i \} \subset \{ p _i \} $ \cite{Kro,Nas08}. Hence, a singular  hyperbolic monopole  can be lifted to a smooth instanton on an appropriately chosen \hgh~space $M$.
Second, the asymptotic conditions on $\Lambda$ ensure that the field strength $F$ of (\ref{inst}) satisfies $ \| F \| ^2 = \mathcal{O} (\exp(-2r) ) $ for large $r$, hence the instanton action on $M$ is finite.

%Finally, as for singular monopoles the monopole action diverges while the action of the corresponding smooth instanton is finite, no simple relation between the instanton and monopole number can be expected to hold.
\subsection{The boundary data of a hyperbolic monopole}
\label{bdata} 
Hyperbolic monopoles are completely determined by the connection $\mathcal{A}_\infty$ on the 
boundary of $H^3$ \cite{BA90,MNS03}.
We can show this explicitly for solutions of the Helmholtz equation, which can be written in the form
\begin{equation}
\label{series}
 \Lambda (r, \theta , \phi )=\sum_{k=0}^{\infty}c_k(\theta,\phi)\,\mathrm{e}^{-(2k+1)r}.
\end{equation}
According to 
\eqref{phia2}, \eqref{asb}, the boundary connection is
\begin{equation}
 \mathcal{A}_\infty=\frac{\mathbf{i} }{2}\left(\cos\theta\,\dd\phi+\ast_{S^2}\dd f_0\right)
\end{equation}
and will completely determine the monopole as long as the fields $\Phi$, $\mathcal{A} $ depend only on the derivatives of $f_0$.
Imposing the Helmholtz equation order by order in (\ref{series})  results in the recursion relation
\begin{equation}
 c_{k+1}=c_k-\frac{1}{(k+1)^2}\sum_{j=0}^k\triangle_{S^2}c_j
\end{equation}
with $c_0=\mathrm{e}^{f_0}$. Thus $\log\Lambda$ can be written in the form \eqref{asb} with  $f_k$, $k\geq1$, depending 
only on derivatives of $f_0$, and the result follows.

The asymptotic expansion of the Higgs field of our hyperbolic monopoles thus only contain terms which decay exponentially with respect to hyperbolic distance.  This is in contrast with what happens for the Higgs field of a  Euclidean monopole, which has both an algebraic tail and an exponential 
falloff near the core.  

Taking $c_0=\mathrm{const}$, we obtain the solution  $\Lambda\propto(\sinh r)^{-1}$, which we will discuss in the next section.  
For a typical solution of the form \eqref{akakaka}, the expansion \eqref{series} breaks down at $\theta=0,\pi$, where a more 
careful analysis is required.  However, for 
the simple solution $f_0=-2\log(\sin\theta)$ we can restrict to the plane $\theta=\pi/2$ to find 
$\Lambda_{\theta=\pi/2}=(2\cosh r)^{-1}$ and $(\partial_\theta\Lambda)_{\theta=\pi/2}=0$, which gives the Higgs field of 
the smooth charge $1$ monopole, $\|\Phi\|=\tfrac{1}{2}\tanh r$.

\subsection{Superposing singular monopoles}
\label{superposition}
The simplest solution of (\ref{helhyp}) with the asymptotic behaviour specified by (\ref{asb}) is
\begin{equation}
\Lambda
= \frac{1 }{\sinh r},
\end{equation}
It corresponds to the Abelian monopole 
\begin{align} 
\label{lambdapole}
\mathcal{A} 
&= \frac{\mathbf{i} }{2} \cos \theta \, \mathrm{d} \phi ,\qquad 
\Phi=- \frac{\mathbf{i} }{2}\coth r,
\end{align} 
which has a singularity at $r =0 $ with Abelian charge  $\ell =1 $ and total charge $ Q =1 $.

\begin{figure}
\centering
\includegraphics[width=0.6\linewidth]{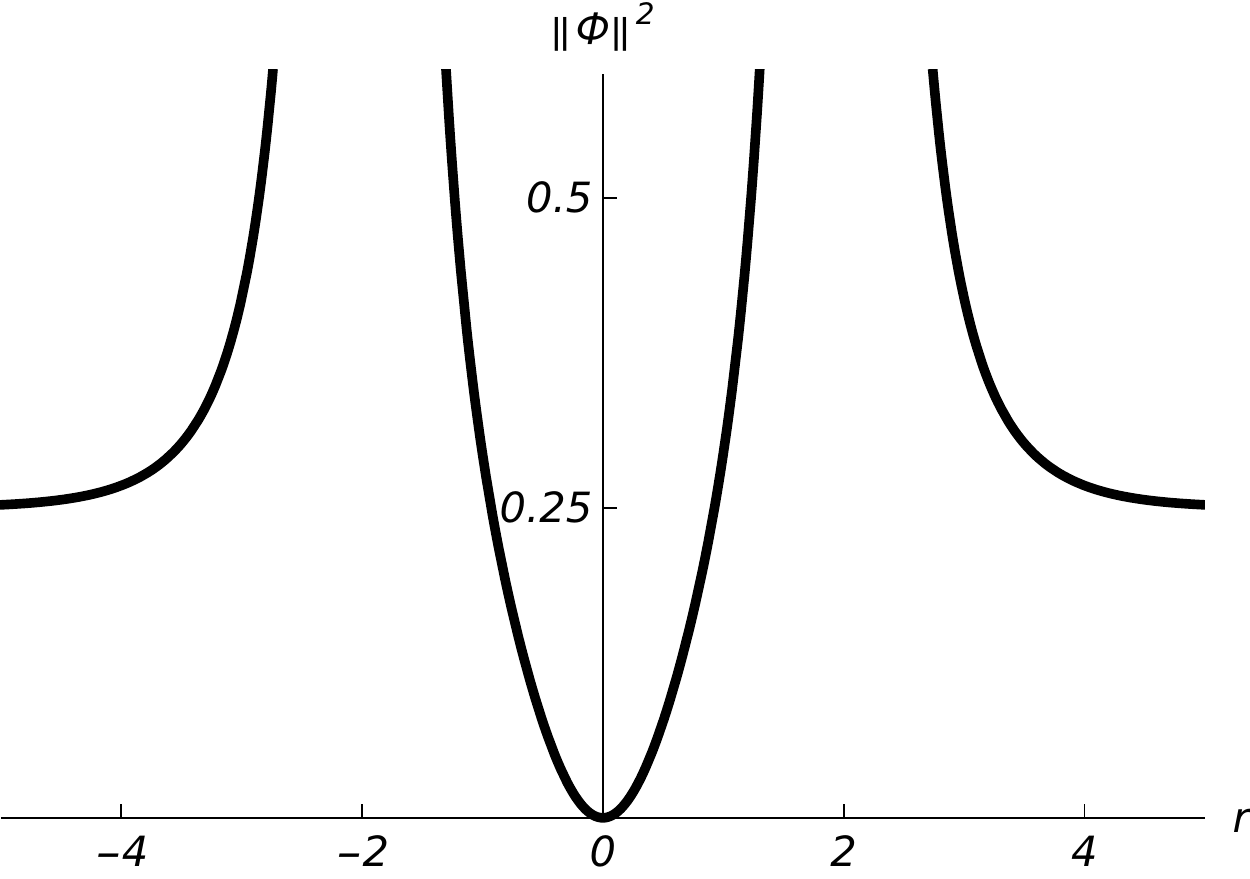}
\caption{The Higgs field along the polar axis ($\theta=0,\pi$) of a monopole constructed by superposing poles of $\Lambda$ at 
$(r,\theta)=(2,0)$ and at $(r, \theta )=(2,\pi)$.}\label{ssmpic}
\end{figure}

Further solutions of (\ref{helhyp}) are obtained by superposing the Green functions (\ref{lambdapole}) at distinct points 
$\{p_i\}$,
\begin{equation}
 \Lambda
 = \sum_{i=1}^n\frac{1}{\sinh(d ^H (p,p_i))}.
\end{equation}
The resulting monopole configuration has Abelian charge $\ell=n$ and total charge $Q=1$.  From (\ref{klq}) it has non-Abelian 
charge $k=n-1$. The non-Abelian charge is reflected in the number of zeros of the Higgs field $\|\Phi\|$.  For example, the 
Higgs field for $n=2$ has two poles with a zero at the midpoint, see Figure \ref{ssmpic}.  Notice that the 
configuration in which all the poles and zeros are placed on the $\theta=0$ axis is axially symmetric.  This should be 
contrasted with the case of strings of non-Abelian monopoles, for which there is a breaking of axial symmetry \cite{Mal}.

\subsection{Generating solutions of the Helmholtz equation}
\label{Backlund}
Let $(M, g) $ be a scalar flat conformally \hgh~space,
\begin{equation}
\label{confg}
g
\, =\,  \Omega  ^2 \, g _0 
\,=\, \Omega  ^2 \left[ V g_{H^3}+V^{-1}(\dd\psi+\alpha)^2 \right].
\end{equation} 
As shown in Section \ref{hyphel}, $g$ is conformally flat if $\Omega$ satisfies the Helmholtz equation,
\begin{equation}
\label{omhel} 
\triangle _{ H ^3 } \Omega + \Omega =0.
\end{equation} 
% \gf{Given a $\psi$-independent harmonic function $\rho$ on $M $  we can generate another solution of the Helmholtz equation.  In fact, set $\Lambda =\rho \, \Omega $.  Then
% \begin{equation} \begin{split}
% 0 &
% =\triangle _{M } \rho 
% = * _{ M } \mathrm{d}  * _{ M }\mathrm{d}  \left( \Lambda / \Omega \right) 
% = - \frac{2}{\Omega ^2 } \langle \mathrm{d} \Omega , \mathrm{d} \Lambda \rangle _{ M } + \frac{1}{\Omega } \triangle _{ M}\Lambda - \frac{\Lambda }{\Omega ^2 } \triangle _{ M }\Omega +  \frac{2\Lambda }{\Omega ^3 } \left | \mathrm{d} \Omega \right | _{ M }^2 .
% \end{split}
% \end{equation} 
% For $\psi$-independent functions $h , k $ on $M $ it holds
% Hence, making  also use of (\ref{omhel}),  we see that $\Lambda$ is also a solution of the Helmholtz equation (\ref{hel}).}
The Laplacian of a $\psi$-independent function $h$ on $M $ is
\begin{equation}
\label{hkformulae}
 \triangle _{ M  } h  
=\frac{1}{\Omega ^2 V  } \left( \triangle _{ H ^3 } h  + \frac{2}{\Omega } \langle \mathrm{d} h  , \mathrm{d} \Omega \rangle _{ H ^3 } \right),
\end{equation}
where $\langle\dd h,\dd\Omega\rangle_{H^3}=\ast_{H^3}(\dd h\wedge\ast_{H^3}\dd\Omega)$.  To find new monopole solutions we 
replace $\Omega$ in \eqref{confg} by $\Lambda=\rho\,\Omega$ and impose that $\Lambda^2g_0$ is scalar flat, $\triangle_{H^3}\Lambda+\Lambda=0$.  
Using \eqref{omhel} one can then see that $\triangle_M\rho=0$.  This procedure allows us to obtain a new solution of the 
Helmholtz equation starting from a simpler solution together with a harmonic function.

We illustrate the procedure by constructing a smooth non-Abelian $k=1$ hyperbolic monopole starting from a Yang-Mills instanton on  Eguchi-Hanson (EH) gravitational 
instanton.  The metric on EH space is often expressed in Euclidean Gibbons-Hawking form,
\begin{equation}
\label{ehgh} 
g _{ \mathrm{EH} }
= \mathcal{V} \left(\dd\tr^2+\tr^2\left(\dd\tth^2+\sin^2\tth\,\dd\psi^2\right)\right) + \mathcal{V} ^{-1} (2\,\mathrm{d}\phi + \aa )^2,
\end{equation} 
where $\dd\mathcal{V}=\ast_{E^3}\dd a$.
Here  $u\in[0, \infty )$, $\chi \in[0, \pi] $, $\psi\in[0, 2 \pi )$, $\phi \in[0,2 \pi )$. The Abelian monopole $(\mathcal{V},a)$ has two poles,
\begin{equation}
\label{euclv} 
\mathcal{V} 
= \frac{1}{\tr_+}+\frac{1}{\tr_-},\qquad \tr_\pm^2=\tr^2+\tr_0^2\pm2\tr\tr_0\cos\tth,
\end{equation} 
where $\tr _0 $ is an arbitrary constant.
%\red{$\ast\ast\ast$ implying that $\phi\in[0,2\pi)$. $\ast\ast\ast$}

It is a remarkable fact that EH space is  conformal to a hyperbolic Gibbons-Hawking metric \cite{ALB13}.  This is 
achieved by the following coordinate transformation (originating from \cite{Pra79})
\begin{equation}
\label{ehghcoords}
 \tr^2=u_0^2\left(\frac{1}{\sinh^2r}+\cos^2\theta\right),\qquad\tan\tth=\frac{\tan\theta}{\cosh r}.
\end{equation}
In these coordinates the EH metric reads
\begin{equation}
\label{hgheh}
 g_{\mathrm{EH}}=\frac{2\,u_0}{\sinh^2r}\left[V(\dd r^2+\sinh^2r(\dd\theta^2+\sin^2\theta\,\dd\phi^2))+V^{-1}\left(\dd\psi+\cos\theta\,\dd\phi\right)^2\right],
\end{equation}
with $V=(\tanh r)^{-1}$.  We remark that when transforming between \eqref{ehgh} and \eqref{hgheh}, the roles of 
$\phi$ and $\psi$ are interchanged.  The metric inside square brackets  has the form (\ref{HGH}) with $V$ given by (\ref{vvv}), a single pole at the origin and  $\beta =2 $. However, the range of $\psi$ in (\ref{hgheh})  is $[0, 2 \pi )$, while the metric (\ref{HGH}) of an \hgh~space has $\psi \in[0, 4 \pi )$. Therefore, a $\beta  =2 $ single pole \hgh~space is conformally equivalent to a branched double cover of the EH space, in which large $r$ hypersurfaces have the topology of $SU(2) $ rather than that of $SO (3) $. 

Let us now look for monopole solutions.  As EH space is scalar flat, we immediately recover the Abelian monopole \eqref{lambdapole} from the conformal factor 
$\Omega\propto(\sinh r)^{-1}$.  
In order to find other monopoles we need harmonic functions on EH space. For this purpose it is convenient to work with the metric in the form \eqref{ehgh}.  A $\phi $-independent function  $\rho$ on EH space satisfies 
$\triangle_{EH}\rho=\mathcal{V}^{-1}\triangle_{E^3}\rho$, therefore a $\phi$-independent  harmonic function on EH space is simply a harmonic  function on $E^3$.  In particular, $\rho=\mathcal{V}$ is a suitable  function.

By expressing $\mathcal{V}$ in $(r,\theta)$ coordinates we get a new solution of the Helmholtz equation,
\begin{equation}
 \Lambda=\mathcal{V} \,\Omega=\frac{4\,\cosh r}{1+\sinh^2r\sin^2\theta}.
\end{equation}
Equation \eqref{phia} shows that this is the smooth spherically symmetric $k=1$ non-Abelian monopole, for which 
\begin{equation} 
\label{lambdasmooth} 
\|\Phi\|=\tfrac{1}{2}\tanh r.
\end{equation}
  Furthermore,  the manifold with the conformally related metric 
$(\Lambda/\Omega)^2g_{\mathrm{EH}}$ has finite volume.

\begin{figure}
\centering
\includegraphics[width=0.6\linewidth]{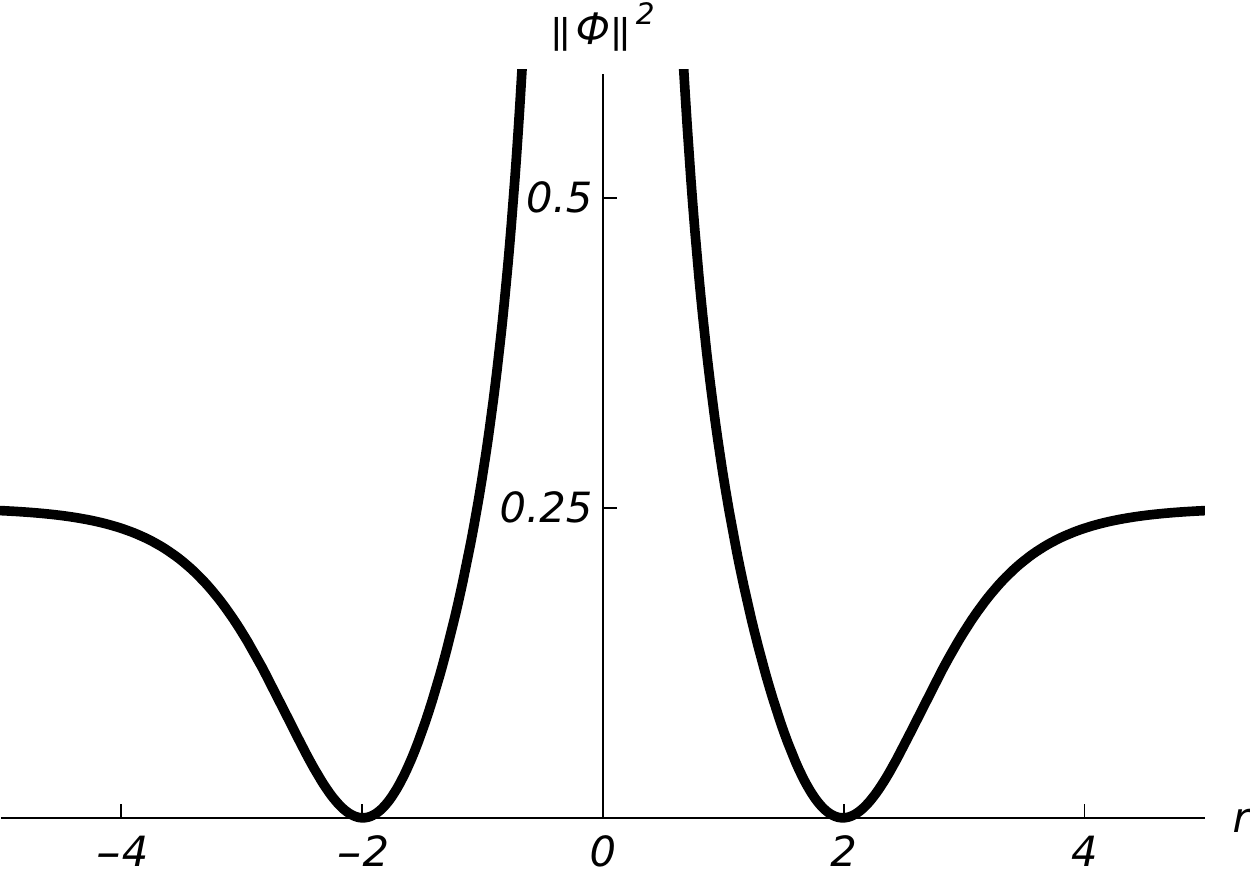}
\caption{The Higgs field along the polar axis ($\theta=0,\pi$) of a monopole constructed from the harmonic function  
$\rho=\lambda+\mathcal{V}$, with $\mathcal{V} $ given by  (\ref{euclv}). Here $\lambda=200$.}\label{bacpic}
\end{figure}

Another singular monopole is obtained by adding a constant $\lambda$ to the harmonic function $\mathcal{V} $. The resulting configuration 
has $\ell =1 $, $Q =-1 $, hence $k = 2 $. See Figure \ref{bacpic} for a plot of the squared norm of the Higgs field. The singularity is located at $r =0 $, and aligned with two equally spaced zeros 
whose separation depends on the value of $\lambda$.

% Euclidean 4-space is conformally equivalent to $H ^3 \times S ^1 $. Let $\psi$ be an angular coordinate on the $S ^1 $ factor and write
%\begin{equation}
%g _{ E ^4 } =\Omega ^2 \left( g _{  H ^3 } + \mathrm{d} \psi ^2 \right) .
%\end{equation} 
%As $E ^4 $ is scalar flat and $g _{  H ^3 } + \mathrm{d} \psi ^2 $ is a special case of an \hgh~space corresponding to $V=1$, 
%
%Let $ \rho $ be a  harmonic function on $E ^4 $ and set $\Lambda =\rho \, \Omega $.  Then
%\begin{equation} \begin{split}
%0 &
%=\triangle _{ E ^4 } \rho 
%= * _{ E ^4 } \mathrm{d}  * _{ E ^4 }\mathrm{d}  \left( \Lambda / \Omega \right) 
%= - \frac{2}{\Omega ^2 } \langle \mathrm{d} \Omega , \mathrm{d} \Lambda \rangle _{ E ^4 } + \frac{1}{\Omega } \triangle _{ E ^4 }\Lambda - \frac{\Lambda }{\Omega ^2 } \triangle _{ E ^4 }\Omega +  \frac{2\Lambda }{\Omega ^3 } \left | \mathrm{d} \Omega \right | _{ E ^4 }^2 .
%\end{split}
%\end{equation} 
%For functions $h , k $ on $E ^4 $ which are independent of $\psi$ it holds
%\begin{equation}
%\begin{split} 
%\langle \mathrm{d} h ,\mathrm{d} k  \rangle _{ E ^4 } &
%=\frac{1}{\Omega ^2 } \langle \mathrm{d} h  , \mathrm{d} k  \rangle _{ H ^3 },\\
% \triangle _{ E ^4 } h  &
%=\frac{1}{\Omega ^2 } \left( \triangle _{ H ^3 } h  + \frac{2}{\Omega } \langle \mathrm{d} h  , \mathrm{d} \Omega \rangle _{ H ^3 } \right).
%\end{split} 
%\end{equation}
%Hence, making  also use of (\ref{omhel}),  it follows that $\Lambda$ is also a solution of the Helmholtz equation (\ref{hel}) on $H ^3 $.

\subsection{JNR equivalence}
\label{jne} 
A large family of Yang-Mills instantons on $E^4$ can be constructed from the JNR ansatz \cite{JNR77}.  The prescription takes a harmonic 
function $\rho$ on $E^4$ of the form
\begin{equation}
\label{JNRfunction}
 \rho=\sum_{i=0}^N \lambda _i \,  G _{ p _i } ^{ E ^4 },
\end{equation}
where $G _{ p _i } ^{ E ^4 } (p) =|p - p_i|^{-2}$, with $ |\cdot |$  the Euclidean distance.
A charge $N$ instanton is then constructed as
\begin{equation}
\label{JNRA}
 A = - \frac{1}{2}\,\sigma_{\mu\nu}\, \partial^\nu\log\rho\,\,  \dd x^\mu,
\end{equation}
where the tensor $\sigma$ is given in terms of the unit quaternions and the anti-self-dual 't~Hooft matrices \eqref{asdth} by 
$\sigma=\mathbf{i}\,\bar{\eta}_1+\mathbf{j}\,\bar{\eta}_2+\mathbf{k}\,\bar{\eta}_3$.

 In order to obtain a hyperbolic 
monopole, one makes use of the conformal equivalence between $E^4$ and $H^3\times S^1$,
\begin{equation}
\label{e4hgh}
 g_{E^4}=z^2\left[g_{H^3}+\dd\psi^2\right]=z^2\left(\frac{\dd x^2+\dd y^2+\dd z^2}{z^2}+\dd\psi^2\right).
\end{equation}
Circle invariance is ensured by placing all the poles $p_i$ of $\rho$  on the fixed plane of the 
$\partial / \partial \psi$ action,  a $2$-plane in $E^4$ which maps to the boundary of $H^3$.  In a circle invariant gauge write $ A = \mathcal{A} + \Phi \, \mathrm{d} \psi $.  Then $( \Phi, \mathcal{A}  )$ is a hyperbolic monopole. The Higgs field has squared norm \cite{BCS15}
\begin{equation}
 \|\Phi\|^2=\frac{z^2}{4\rho^2}\left(\left(\partial_x\rho\right)^2+\left(\partial_y\rho\right)^2+\left(\frac{\rho}{z}+\partial_z\rho\right)^2\right)=-\frac{1}{4}\left(1+\triangle_{H^3}\log(z\rho)\right),
\end{equation}
where for the second equality we have made use of the equations
\begin{IEEEeqnarray}{rcl}
 \triangle_{H^3}\rho&\,=\,&z^2(\partial_x^2+\partial_y^2+\partial_z^2)\rho-z\partial_z\rho,\\
 z^2\triangle_{E^4}\rho&\,=\,&z^2(\partial_x^2+\partial_y^2+\partial_z^2)\rho+z\partial_z\rho=0.
\end{IEEEeqnarray}

Note that $H ^3 \times S ^1 $ is a special case of an \hgh~space with $V =1 $, $ \alpha = 0 $. Since $E ^4 $ is scalar flat, we recover the JNR construction of monopoles by applying the method of Section \ref{Backlund}:~take the conformal factor $ \Omega =z $ and a  harmonic function $\rho $ on $E ^4 $, then  $\Lambda =z \rho $ generates a hyperbolic monopole via equations (\ref{toph}), (\ref{topgp}).

% The conformal equivalence between $E^4$ and $H^3\times S^1$
% \begin{equation}
% \label{e4hgh}
%  g_{E^4}=z^2\left[g_{H^3}+\dd\psi^2\right]=\dd x^2+\dd y^2+\dd z^2+z^2\dd\psi^2,
% \end{equation}
% is a special case of the construction of Section \ref{Backlund}, corresponding to $V=1$, $\alpha=0$ and half-space model 
% coordinates on $H^3$ so that $\Omega =z $.  Hence we can  generate monopole solutions by taking $\Lambda=z\rho$, with $\rho$ a 
% $\psi$-independent harmonic function on $E^4$.  For the particular choice
% \begin{equation}
% \rho 
% = \sum _{ i } \lambda _i \, G ^{  E ^4 } _{p _i },
% \end{equation} 
% where $G ^{ E ^4 } _{p _i} $ is the Green's function on $E ^4 $ centred at $p _i $, we recover the JNR construction. Circle 
% invariance is ensured by taking all the $ p _i $s to lie in the 2-plane fixed under the circle action, which is mapped to the 
% boundary of $H ^3 $ under the conformal isometry (\ref{e4hgh}).  The corresponding monopole has $p =1/2 $.

%As we have just seen, any solution of the Helmholtz equation corresponds to a harmonic function on $E ^4 $, and viceversa. 
We expect that all the solutions of the Helmholtz equation giving  smooth hyperbolic monopoles correspond to harmonic functions on $E ^4 $ which are of JNR type.  To translate from JNR data to solutions of the hyperbolic Helmholtz equation in hyperboloid model coordinates we make use of the coordinate transformation
\begin{alignat}{2}
\nonumber
x &=\frac{\sinh r \sin \theta \cos \phi }{\cosh r - \sinh r \cos \theta } &\qquad\qquad   \tan \phi &=\frac{y}{x},\\ \label{ajdHVJHaaaasd}  
y &=\frac{\sinh r \sin \theta \sin  \phi }{\cosh r - \sinh r \cos \theta } & \tan ^2  \theta  &= \frac{4( x ^2 + y ^2 ) }{ (x ^2 +  y ^2 + z ^2 -1 )^2 }, \\ \nonumber
z &=\frac{1 }{\cosh r - \sinh r \cos \theta } &\cosh r &=   \frac{ x ^2 + y ^2 + z ^2 + 1  }{2z } .
\end{alignat}

Singular hyperbolic monopoles can also be described in terms of JNR data.
Using (\ref{ajdHVJHaaaasd}) we find that the JNR  function corresponding to the singular monopole with $\Lambda=(2\sinh r)^{-1}$ is
\begin{equation}
 \rho
 = \frac{\Lambda }{z}
 =\frac{1}{\sqrt{\left(1+x^2+y^2+z^2\right)^2-4z^2}}.
\end{equation}
Note that the function $\rho$  is singular at the monopole location, $(x,y,z)=(0,0,1)$.  A monopole singularity may thus be interpreted as a JNR pole which has been 
moved from the boundary of hyperbolic space to the interior.  This interpretation of singular monopoles explains our 
observation in Section \ref{superposition} that by superposing two Abelian monopoles the Higgs field  acquires a zero.  Separating the poles 
causes the profile of the Higgs field near the zero to approach that of the smooth unit charge non-Abelian monopole.  In general, a 
JNR configuration consisting of $n>0$ poles on the boundary of $H^3$ and $\ell\geq0$ poles in the interior has Abelian charge 
$\ell$, non-Abelian charge $k=n+\ell-1$ and total charge $Q=1-n$ independent of $\ell$.

\subsection{Higher mass monopoles}
\label{hmm} 
Dimensional reduction of instantons also gives monopoles of mass $m>\tfrac{1}{2}$.  Take the axially 
symmetric instantons on Eguchi-Hanson space described by Boutaleb-Joutei {\it et al.} \cite{BCC80}.  In the coordinates 
\eqref{hgheh} their solution reads
\begin{equation}
\label{BCCphia}
 A_r=0,\qquad A_\theta=-D\,\frac{\mathbf{k}}{2},\qquad A_\phi=G\cos\theta\,\frac{\mathbf{i}}{2}-D\sin\theta\,\frac{\mathbf{j}}{2},\qquad A_\psi=(G-1)\,\frac{\mathbf{i}}{2},
\end{equation}
where
\begin{equation}
 D=\frac{\alpha\sinh r}{\sinh(\alpha r)},\qquad G=\frac{\alpha\tanh r}{\tanh(\alpha r)}.
\end{equation}

This instanton is manifestly circle invariant so we 
can reduce it by a circle action of weight $\tfrac{1}{2}$ to get a hyperbolic monopole.  As far as we know this has not been 
noticed before.  The monopole Higgs field $\Phi=VA_\psi$, where $V=\coth r$, is given by
\begin{equation}
\label{phimass}
 \|\Phi\|=\frac{1}{2}(\alpha\coth(\alpha r)-\coth r)
\end{equation}
and has mass $\tfrac{1}{2}(\alpha-1)$.  This monopole  arises either as a higher weight 
reduction of an axially symmetric Euclidean instanton \cite{Nas86} or, as we have just shown, as a weight $\tfrac{1}{2}$ reduction of an instanton on 
Eguchi-Hanson space.

The modified Helmholtz equation
\begin{equation}
 \triangle_{H^3}\Lambda+K\Lambda=0, \quad K<1
\end{equation}
 has solutions of the form
\begin{equation}
 \Lambda=\frac{\sinh(\sqrt{1-K}\,r)}{\sinh r}.
\end{equation}
It is remarkable that if we use \eqref{philambda} to compute $\| \Phi \| $ 
%\begin{equation}
% \|\Phi\|=\frac{1}{2}\,|\dd\log\Lambda|_{H^3},
%\end{equation}
we recover \eqref{phimass} with $\alpha=\sqrt{1-K}$.  While the pair $(\Phi, \mathcal{A})$ obtained using \eqref{phia}, 
\eqref{phia2} is not a solution of the Bogomolny equations, this suggests that our construction can be extended to 
$m>\tfrac{1}{2}$.

\section{A family of conformally Einstein manifolds}
\label{emono} 
We still have not made use of the second method of Theorem \ref{ahs} which will give us a family of monopoles which, as far as we know, has not been discussed before.
It has been proved \cite{PT91} that an \hgh~space is conformally Einstein if and only if $V$ is spherically symmetric. 
Therefore we  take 
\begin{equation}
\label{nhgtyru} 
V = \frac{2}{\beta } + \frac{2}{\mathrm{e} ^{ 2r} - 1},
\end{equation} 
so that $\alpha = \cos \theta \, \mathrm{d} \phi $.\footnote{Note that with our conventions $V$ in (\ref{nhgtyru}) is twice that appearing in \cite{ALB13}. Consequently, our conformal factor is one half of theirs. } 
The metric 
\begin{equation}
\label{metric} 
g = \frac{2}{\beta ((2 - \beta ) \cosh r  + \beta \sinh r) ^2 } \,\left(V g_{H^3}+V^{-1}(\dd\psi+\alpha)^2\right) , \quad \beta \in(0,2] 
\end{equation} 
is Einstein with  constant 
$\tfrac{3}{2} \beta ^2  (2 - \beta )
$ \cite{ALB13}.
The case $\beta =2 $ corresponds to (a branched double cover of) the Eguchi-Hanson space discussed previously. For $\beta =1 $ one obtains the Fubini-Study metric on $ \mathbb{C}  P ^2 $. With the rescaling $ r \rightarrow r / \beta $ the pointwise limit for $\beta \rightarrow 0 $ of  (\ref{metric}) is the Taub-NUT metric. 

To get a self-dual instanton we need to project the spin connection $\omega$  onto $S ^+ (M) $,
\begin{equation}
P _{ + }(\omega)  
=\frac{1}{4} ( \eta _a ) _{ i  j} \, \omega _{ i j} \,  \eta _a
= \frac{1}{2} \Big[ 
(\omega _{ 34 }+ \omega _{ 12 } )\mathbf{i} + 
 (\omega _{ 24 }- \omega _{13 }) \mathbf{j} + 
(\omega _{ 14 }+ \omega _{23 } )\mathbf{k} 
 \Big],
\end{equation} 
where $ \{\eta _i \}  $ are the  self-dual 't Hooft matrices
\begin{equation}
 \eta _1 
= \begin{pmatrix}
 0 &1 &0 &0\\ -1 &0 &0 &0\\0 &0 &0 &1 \\ 0 &0 &-1 &0\end{pmatrix}, \quad 
 \eta _2 
= \begin{pmatrix}
0 &0 &-1 &0\\ 0 &0&0 &1\\ -1 &0 &0 &0\\ 0 &1 &0 &0
\end{pmatrix},\quad 
 \eta _3 =\begin{pmatrix}
0 &0&0&1\\0&0&1&0\\0&-1&0&0\\-1&0&0&0
\end{pmatrix},
\end{equation} 
and we have identified $ \eta _1 = - \mathbf{k} , \eta _2 = - \mathbf{j}, \eta _3 = - \mathbf{i} $.

\begin{figure}
\centering
\includegraphics[width=0.7\linewidth]{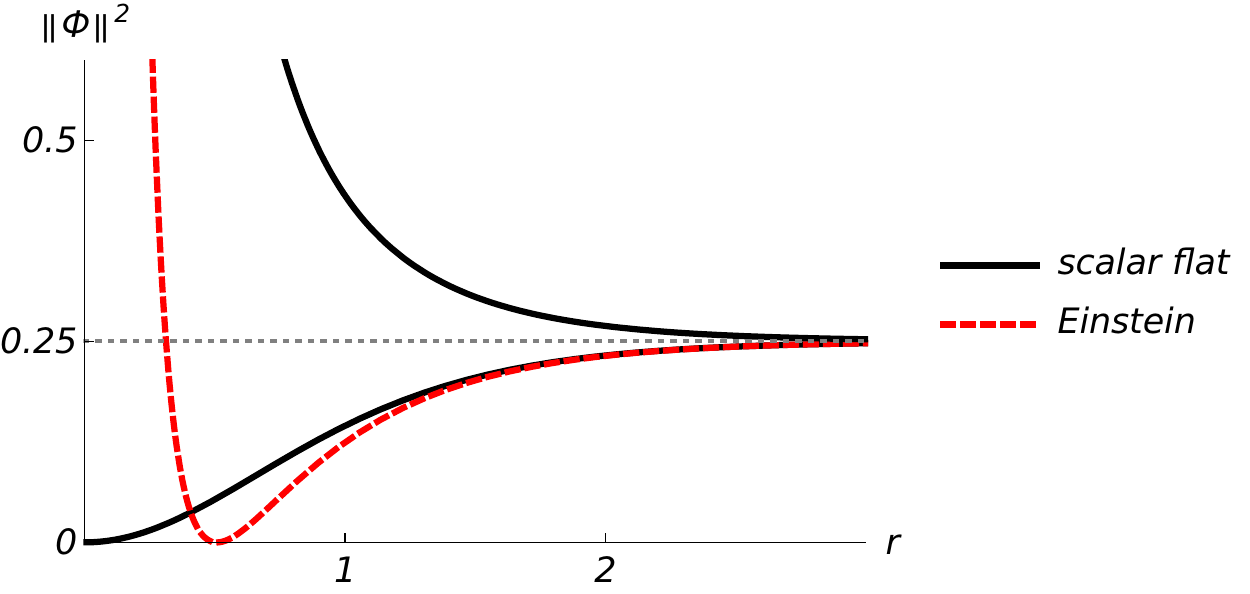}
\caption{Solid curves show the Higgs field profile of the smooth monopole (\ref{lambdasmooth}) and singular monopole 
(\ref{lambdapole})  constructed by conformally rescaling (\ref{HGH}) to be scalar flat. The dashed curve shows the Higgs field 
(\ref{fiei}) for $\beta=1.5$  obtained by rescaling (\ref{HGH}) to be Einstein.  The sign of $Q$ depends on whether the 
asymptotic value $m=\tfrac{1}{2}$ is approached from above ($Q>0$) or from below ($Q<0$).}\label{empic}
\end{figure}

The corresponding instanton has gauge potential $A = \mathcal{A} + \Phi ( \mathrm{d} \psi + \cos \theta \, \mathrm{d} \phi  )/ V  $. The fields $\phi$, $\mathcal{A} $ satisfy the Bogomolny equations (\ref{bogo}) and are given by
\begin{align}
\mathcal{A} &
= \frac{\mathbf{i} }{2} \cos \theta \, \mathrm{d} \phi 
+ \frac{2 \, ( \beta -1 )  \mathrm{e} ^{ 2r } \sinh r}{\mathrm{e} ^{ 4r } - (\beta -1 )^2  }\left(\mathbf{j}\sin \theta \, \mathrm{d} \phi-\mathbf{k}\,\mathrm{d} \theta\right),\\
\label{fiei} 
\Phi &
=\frac{\mathbf{i} }{4} (1 - \coth r )\left( \frac{ (\beta -1 )^2 (1-3 \mathrm{e} ^{ 2r }) - \mathrm{e} ^{ 6r }  (1 -3  \mathrm{e} ^{- 2r }) }{\mathrm{e} ^{ 4r } - (\beta -1 )^2  } \right) .
\end{align} 
Note that $\Phi$, $\mathcal{A}$ are  invariant under the change $\beta \rightarrow 2 - \beta$, coupled to a gauge transfomation 
by $\mathbf{i}$. The monopole has mass $\tfrac{1}{2}$ and total charge $Q = - 1$.  The value $\beta =2 $ gives the  smooth spherically symmetric $k = 1$ 
monopole with $\| \Phi\| = \tfrac{1}{2}\tanh r $. For $\beta \neq 2 $ there is a singularity at $r =0 $ with Abelian weight 
$\ell =1 $ and a $2$-sphere worth of zeros given by $r=r_0\in(0,\tfrac{1}{2}\log(3)]$ as $2-\beta$ varies in $(0,1]$. The Higgs field profile is shown in Figure \ref{empic}. Note that $\Phi$, $\mathcal{A}$ Abelianise  for both small and large values of $r$. Plotting 
the energy density shows no special features at $r=r_0$.  In the limiting case $\beta=1$ the fields Abelianise for all values of $r$  and we have 
$\|\Phi\| =\tfrac{1}{2}|\coth r -2| $.  This family has charge $Q$ opposite to that of the singular monopole $\|\Phi\|=\tfrac{1}{2}\coth r$ we 
encountered before. From \eqref{klq}, we expect the $2$-sphere of zeros to contribute a non-Abelian charge 
$k =2 $.

\section{ Conclusions}
In this work we have shown how a class of smooth and singular hyperbolic monopoles of mass $m=\tfrac{1}{2}$ can be 
expressed in terms of solutions of the hyperbolic Helmholtz equation.  All the solutions we found have an equivalent 
description in terms of JNR data.  However, thinking in terms of the Helmholtz equation presents a number of advantages.

First, our construction is entirely coordinate free, see equations (\ref{toph}) and (\ref{topgp}), while the JNR construction is adapted to the upper half space model of 
hyperbolic space.

Second, we relate \emph{singular} hyperbolic monopoles to \emph{smooth} instantons on a scalar flat $4$-manifold which is conformally equivalent to a hyperbolic 
Gibbons-Hawking space. The JNR data giving the same monopoles describes  instantons on $E^4$ which are singular along circles.  In 
this sense we have illustrated how a conformally Gibbons-Hawking geometry, which is commonly seen as encoding an Abelian 
monopole, also encodes non-Abelian monopoles.

Third, our approach shows in a very explicit fashion, albeit in a special case, how a hyperbolic monopole can be reconstructed 
from its asymptotic data.

Fourth, we provide a physical interpretation of the poles in the JNR ansatz as singular hyperbolic monopoles.  Separating two such 
poles gives rise to a non-Abelian monopole between them.  This should be contrasted with the related Euclidean instanton, for 
which there is no direct physical interpretation of the JNR poles.

Interestingly, the Helmholtz equation also arises in Prasad's generalisation of the Atiyah-Ward ansatz for Euclidean monopoles \cite{Pra81}.

%A final minor commodity is that the instanton obtained out of the Helmholtz equation is already in a circle invariant gauge, so that the monopole can be read off directly.

The manifold $\mathbb{C}  P ^2 $ is conformally \hgh~and Einstein, so it is natural to ask whether circle invariant 
instantons on this space can be reduced to hyperbolic monopoles.  Instantons with instanton number $1$ are studied in \cite{Gro90} 
and there is a $3$-parameter family which is invariant under a circle action.  However, this is not the circle action which 
reduces $\mathbb{C}  P ^2 $ to a  conformally \hgh~space, therefore these instantons do not 
descend to hyperbolic monopoles.

It is natural to ask if our construction can be generalised to $m>\tfrac{1}{2}$. The results described in Section \ref{hmm} 
suggest that, at least for the spherically symmetric case, some more general construction exists. However such an extension is 
not straightforward and we leave it for future work.

\section*{Acknowledgments}
We wish to thank Olaf Lechtenfeld for useful discussions. G.F.~was supported by the DFG Research Training Group 1463.

{\raggedright
\end{document}